\documentclass[runningheads]{llncs}
\usepackage[T1]{fontenc}
\usepackage{graphicx}
\usepackage{amsmath}
\usepackage{booktabs}
\usepackage{float}
\usepackage{hyperref}

\begin{document}

\title{Practice Less, Explain More: LLM-Supported Self-Explanation Improves Explanation Quality on Transfer Problems in Calculus}
\titlerunning{Practice Less, Explain More: LLM-Supported Self-Explanation}

\author{Eason Chen\inst{1}
\and
Xinyi Tang\inst{1}
\and
Yvonne Zhao\inst{1} \and
Meiyi Chen\inst{1} \and
Meryam Elmir\inst{1}
\and
Elizabeth McLaughlin\inst{1}
\and
Mingyu Yuan\inst{1}
\and
Yumo Wang\inst{1}
\and
Shyam Agarwal\inst{1}
\and
Jared Cochrane\inst{1}
\and
Jionghao Lin\inst{2}
\and
Tongshuang Wu\inst{1}
\and
Ken Koedinger\inst{1}
}

\authorrunning{E. Chen et al.}

\institute{Carnegie Mellon University, Pittsburgh, PA, USA 
\\\email{eason.tw.chen@gmail.com}
\and
The University of Hong Kong, Hong Kong, China}

\maketitle

\begin{abstract}
We conducted a between-subjects experiment (N\,=\,92) comparing three conditions in a calculus learning environment: no self-explanation (\textbf{control}), \textbf{menu-based} self-explanation, and \textbf{open-ended} self-explanation with LLM-generated feedback. All conditions showed positive learning gains within a fixed 60-minute practice session, with no significant between-condition differences in post-test performance. On transfer questions, the \textbf{open-ended} condition produced significantly higher-quality explanations than \textbf{control} on ``Not Enough Information'' (NEI) problems ($\beta$\,=\,+11.9 percentage points, $p$\,=\,.030), though the corresponding NEI multiple-choice accuracy advantage was not significant ($p$\,=\,.183). Moreover, across all post-test open-ended explanations, the \textbf{open-ended} condition showed a marginally significant advantage ($\beta$\,=\,+7.3\%, $p$\,=\,.057). These findings suggest that LLM-supported open-ended self-explanation can improve explanation quality on NEI transfer problems, though the evidence is weaker across broader transfer explanation measures. Notably, these effects emerged even though learners in the open-ended condition completed substantially fewer practice problems during the same amount of practice time.

\end{abstract}

\section{Introduction}

Self-explanation, the process of articulating the rationale behind a solution, is one of the most robust learning strategies in educational research \cite{chi1989self,chi1994eliciting,rittlejohnson2017promoting}. Within the ICAP framework \cite{chi2014icap}, self-explanation represents constructive engagement, where learners generate inferences and integrate new knowledge with prior understanding. However, self-explanation demands substantially more time than ordinary practice \cite{chi1994eliciting}, and when total learning time is equated, advantages may diminish. Specifically, Matthews and Rittle-Johnson \cite{matthews2009pursuit} and McEldoon et al.\ \cite{mceldoon2013self} found that self-explanation advantages disappeared when control groups practiced more problems within the same total time; critically, both studies used unprompted verbal explanations without feedback on explanation quality. Previous studies have shown that open-ended explanation quality also varies \cite{aleven2000need,renkl1997learning} widely, limiting effectiveness.

These challenges motivated scaffolded approaches in intelligent tutoring systems (ITS). Aleven and Koedinger \cite{aleven2002effective} developed menu-based self-explanation in a geometry Cognitive Tutor, showing that students who explained their problem-solving steps learned with greater understanding, produced better explanations of solution steps, and were more successful on transfer problems. Their subsequent dialogue-based systems \cite{aleven2001towards,aleven2003tutorial} interpreted natural-language explanations with corrective feedback, achieving equivalent learning with fewer practice problems, but required hundreds of handcrafted rules \cite{aleven2001towards}, limiting scalability. LLMs offer a promising solution: recent work shows they can evaluate educational content with reliability approaching human experts \cite{carpenter2024assessing,lin2024howcani,chen2025praise}, generate effective feedback \cite{kumar2025math}, and support learning at scale \cite{armfield2025avalon,chen2024systematic}.

Transfer, the ability to apply learned knowledge to novel situations, is a central goal of education yet remains challenging to achieve \cite{hajian2019transfer}. Self-explanation may promote transfer by encouraging learners to abstract underlying principles \cite{rittlejohnson2024encouraging}. Despite the extensive literature, very few studies have directly compared open-ended and menu-based self-explanation; to our knowledge, Aleven and Koedinger \cite{aleven2002effective} remains the only such comparison in the ITS literature. To address this gap, we ask whether LLM-supported open-ended self-explanation can improve learners' reasoning on transfer problems relative to lighter-weight alternatives under fixed-time conditions. We investigate this in calculus by comparing three conditions: no self-explanation (\textbf{control}), \textbf{menu-based} self-explanation, and \textbf{open-ended} self-explanation with LLM feedback. We hypothesize that \textbf{open-ended} self-explanation with adaptive feedback will yield stronger performance on open-ended explanations for transfer problems, especially problems requiring learners to recognize when insufficient information is provided.

\section{Method}

\noindent \textbf{Participants.} We recruited 301 U.S.\ participants through Prolific, screening for algebra proficiency and excluding those who had completed three or more college-level math courses. With this filter, 105 failed the screening question, and 104 dropped out, yielding N\,=\,92 with similar attrition across conditions.

\noindent \textbf{Design.} We employed a 3 (Condition) $\times$ 2 (Counterbalance order) between-subjects design. Participants were randomly assigned to: \textbf{Control} (no self-explanation, n\,=\,29), \textbf{Menu-based} (select explanation from options after each problem, n\,=\,35), or \textbf{Open-ended} (write explanation with LLM feedback after each problem, n\,=\,28). Quiz order was counterbalanced (Quiz~A then B, or B then A).

\begin{figure}[H]
    \centering
    \includegraphics[width=0.9\linewidth]{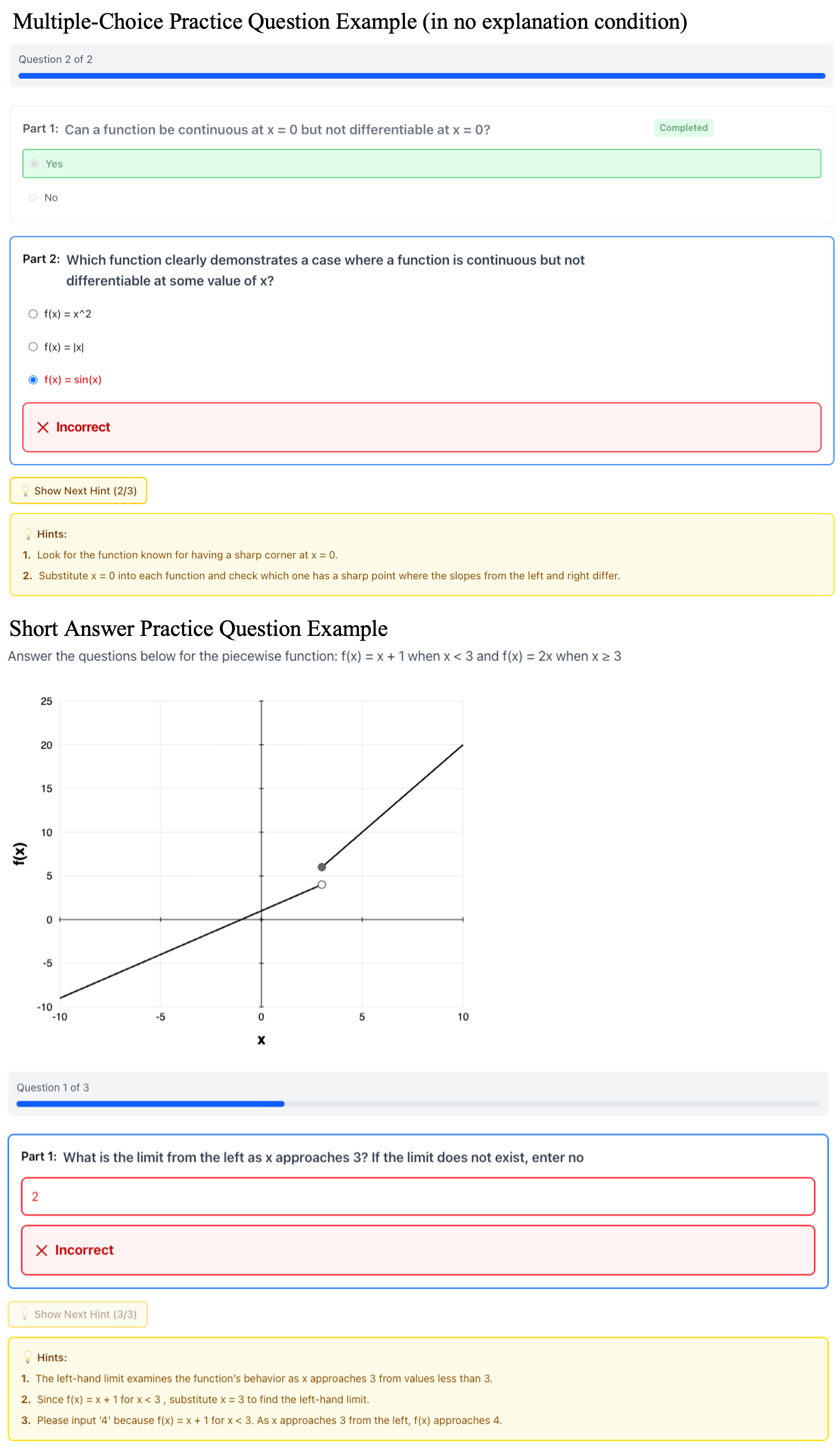}
    \caption{Practice interface (control condition). Top: multiple-choice with correctness feedback and progressive hints. Bottom: short-answer with a piecewise function graph.}
    \label{fig:question_types}
\end{figure}

\noindent \textbf{Materials.} We prepared calculus practice problems targeting procedural and conceptual understanding of limits and derivatives, which were developed by a team including undergraduate and graduate students under expert guidance. Items were balanced across multiple-choice, short-answer, and table-based formats (see Figure~\ref{fig:question_types}). Each problem included three explanation options for the menu-based condition: one correct principle and two plausible misconceptions.

Two equivalent quizzes (Quiz~A and Quiz~B) served as pre/post-tests, each containing problem-solving and transfer questions. Transfer items included ``Not Enough Information'' (\textbf{NEI}) problems: piecewise function questions asking whether the function was differentiable at a given point, where critical slope information was unknown, making the correct answer ``Not enough information.'' Each NEI question had a multiple-choice component (``Yes,'' ``No,'' or ``Not enough information'') and an open-ended explanation component. Transfer items also included ``Enough Information'' (EI) problems, where the necessary information was provided for a definitive answer. Each quiz form contained 4 transfer questions (2 NEI + 2 EI).

\noindent \textbf{System.} All materials were implemented in a NextJS web application deployed on Vercel with PostgreSQL, recording step-level correctness, hints requested, and timestamps. An interactive demo is available at the following link: \url{https://self-explanation-study.vercel.app/preview/questions}.

\noindent \textbf{LLM grading validation.} We designed a four-level rubric (0, 0.3, 0.7, 1.0) for explanation quality, where higher scores reflected more complete and conceptually accurate reasoning. Two human graders evaluated a sample of 160 explanations from a prior pilot study, reaching substantial agreement (quadratic-weighted Cohen's $\kappa$\,=\,0.70); disagreements were resolved by the first author to form the human consensus score. We then evaluated whether an LLM (GPT-5.1), using the same rubric and grading prompt, could approximate this scoring process. The model achieved quadratic-weighted $\kappa$\,=\,0.68 against Grader 1, $\kappa$\,=\,0.66 against Grader 2, and $\kappa$\,=\,0.78 with human consensus score.

\noindent \textbf{Procedure.} The study is approved by the IRB, followed by a pre-test, practice (60-min fixed), and post-test design (average total 115 minutes). No correctness feedback was provided during pre- or post-tests. During practice, all participants received immediate correctness feedback and were required to keep attempting each problem until arriving at the correct answer (Figure~\ref{fig:question_types}). After an incorrect attempt, participants could request a hint; following previous CTAT conventions \cite{aleven2006cognitive}, each hint request was logged as an additional incorrect attempt. When participants reached the 60-minute limit, they completed their current problem and proceeded to the post-test.

\begin{figure}[H]
    \centering
    \includegraphics[width=0.9\linewidth]{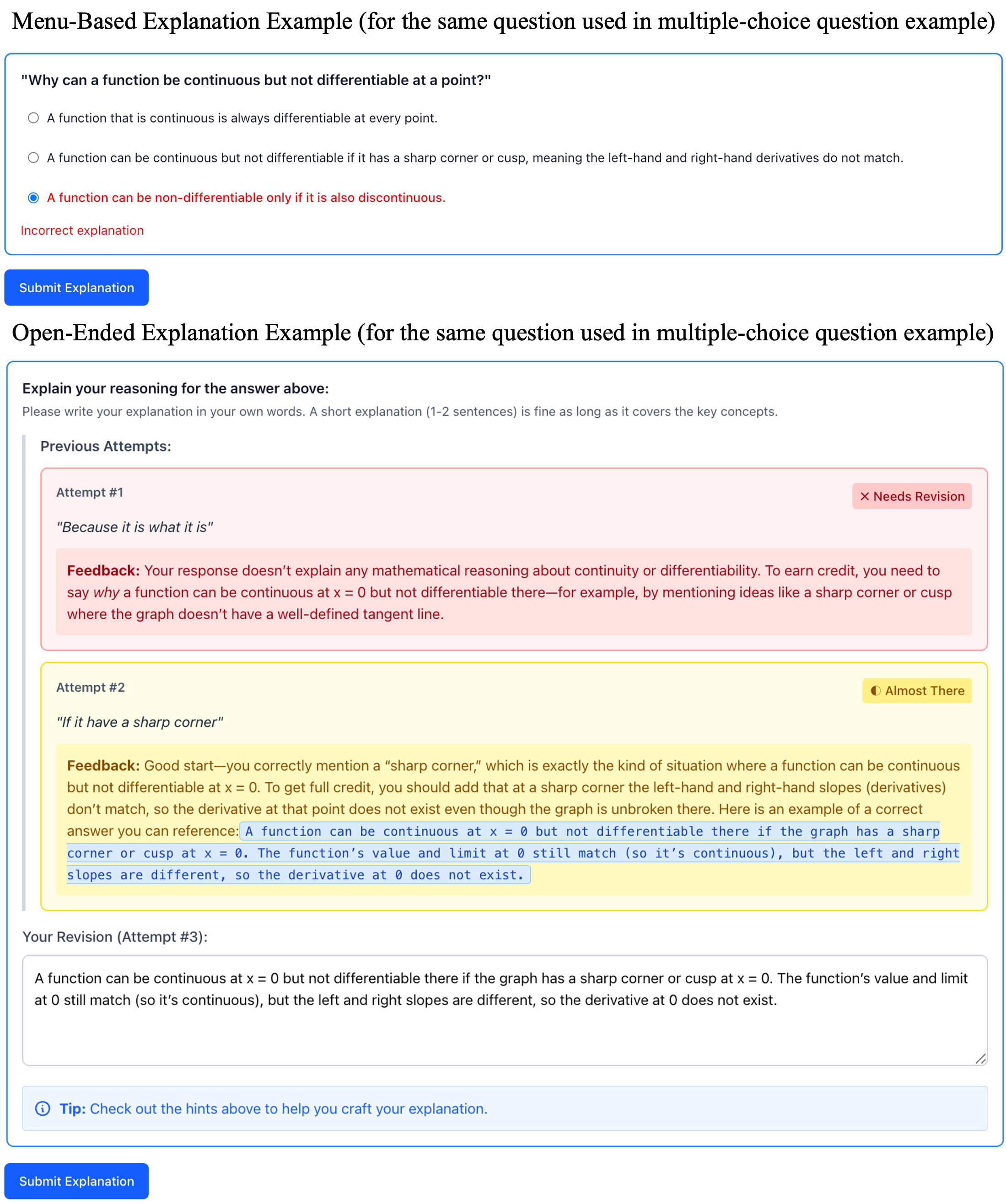}
    \caption{Self-explanation conditions. Top: menu-based, selecting from three options (one correct, two misconceptions). Bottom: open-ended with LLM feedback showing iterative revision (red = needs revision, yellow = almost there) and a reference explanation after two unsuccessful attempts.}
    \label{fig:explanation_styles}
\end{figure}

After solving a problem correctly, participants in the experiment group completed their assigned explanation activity. In the \textbf{menu-based} condition, participants selected the reasoning behind their answer from a list of options (one correct principle and two misconceptions); they received correctness feedback and were required to select the correct option before proceeding. In the \textbf{open-ended} condition, participants wrote free-text explanations evaluated by the LLM with color-coded feedback: red for incorrect (score 0), yellow for partial (0.3--0.7), and green for correct (1.0). After two unsuccessful explanation attempts, the AI provided a reference explanation (Figure~\ref{fig:explanation_styles}).

\noindent \textbf{Outcome measures.} We distinguish three types of post-test outcomes. First, post-test problem-solving accuracy measured performance on problems similar to those practiced. Second, transfer MCQ accuracy measured answer selection on transfer items, including NEI items and non-NEI (EI) items. Third, transfer open-ended explanation quality measured the quality of written explanations on those transfer items, scored by the LLM using the validated rubric described above. Combined OE aggregates open-ended explanation quality across both NEI and non-NEI transfer items.

\noindent \textbf{Analysis.} All analyses used ANCOVA: Outcome $\sim$ Pre-test + Condition + Counterbalance, with Control and AB order as reference groups. We report unstandardized $\beta$ (percentage points) and two-tailed $p$-values.

\section{Results}

Table~\ref{tab:descriptive} presents descriptive statistics by condition. Pre-test scores were similar across conditions (75.2\%--76.0\%), consistent with successful randomization. A notable pattern emerged under the fixed-time design: within the 60-minute practice session, the control condition completed roughly 3.5 times as many problems as the open-ended condition (58.9 vs.\ 16.9), yet both showed similar observed learning gains (+9.3\% vs.\ +9.4\%). This fixed-time contrast is important for interpreting the results: the open-ended condition had far fewer opportunities for ordinary practice, so any advantage on transfer explanations is unlikely to reflect simply solving more problems.

\begin{table}[t]
    \centering
    \caption{Performance by condition ($\pm$ SE).}
    \label{tab:descriptive}
    \begin{tabular}{l c c c c c}
        \toprule
        Condition & n & Pre-test & Post-test & Gain & Problems \\
        \midrule
        Control & 29 & 75.2\% $\pm$ 4.5 & 84.6\% $\pm$ 2.9 & +9.4\% & 58.9 \\
        Menu & 35 & 75.4\% $\pm$ 4.2 & 81.3\% $\pm$ 3.3 & +5.9\% & 43.8 \\
        Open-ended & 28 & 76.0\% $\pm$ 4.6 & 85.2\% $\pm$ 2.7 & +9.3\% & 16.9 \\
        \bottomrule
    \end{tabular}
\end{table}

\begin{table}[t]
    \centering
    \caption{Condition effects (vs.\ Control) across transfer outcomes.}
    \label{tab:all_transfer}
    \begin{tabular}{l c c c c}
        \toprule
        Outcome & Open $\beta$ & $p$ & Menu $\beta$ & $p$ \\
        \midrule
        NEI Transfer OE & +11.9\% & .030* & +4.7\% & .343 \\
        Non-NEI Transfer OE & +6.3\% & .093\textsuperscript{\textdagger} & +1.7\% & .587 \\
        Combined OE & +7.3\% & .057\textsuperscript{\textdagger} & +2.2\% & .513 \\
        \midrule
        NEI Transfer MCQ & +9.3\% & .183 & $-$2.0\% & .730 \\
        Non-NEI Transfer MCQ & +0.4\% & .909 & +3.3\% & .414 \\
        \bottomrule
    \end{tabular}
    
    \smallskip
    \footnotesize{*$p < .05$, \textsuperscript{\textdagger}$p < .10$}
\end{table}

\noindent \textbf{NEI Transfer Open-Ended.} ANCOVA revealed a significant effect: the open-ended condition scored 11.9 percentage points higher than control ($\beta$\,=\,+11.9\%, SE\,=\,5.8\%, $t$\,=\,2.06, $p$\,=\,.030, 95\% CI [1.2\%, 22.6\%], $d$\,=\,0.44). Menu-based did not differ from control ($\beta$\,=\,+4.7\%, $p$\,=\,.343). After controlling for pre-test score (a strong predictor, $\beta$\,=\,0.509, $p$\,$<$\,.001), the model explained 20.4\% of variance (Adj.\ $R^2$\,=\,.204).

\noindent \textbf{Learning gains.} ANCOVA on post-test scores revealed no significant condition effects (menu: $\beta$\,=\,$-$2.4\%, $p$\,=\,.564; open-ended: $\beta$\,=\,+0.8\%, $p$\,=\,.842).

\noindent \textbf{Transfer outcomes summary.} Table~\ref{tab:all_transfer} presents the full pattern. We did not find significant condition differences on ordinary post-test performance or on transfer answer selection (MCQ), including NEI transfer MCQ. The significant effect appeared specifically in the LLM-scored explanations for NEI transfer items. Effects for the open-ended condition were also positive across open-ended explanation measures, with combined post-test open-ended explanations yielding a marginally significant effect ($\beta$\,=\,+7.3\%, $p$\,=\,.057, $d$\,=\,0.38). NEI Transfer MCQ showed a non-significant advantage in the same direction ($\beta$\,=\,+9.3\%, $p$\,=\,.183, $d$\,=\,0.26). Together, these results indicate stronger evidence for improved explanatory reasoning on NEI transfer problems than for broader gains in transfer answer accuracy.

\section{Discussion}

This study examined whether LLM-supported open-ended self-explanation can enhance calculus learning and performance on transfer tasks. Three key findings emerged. First, compared to the menu-based explanation and the no-explanation control group, we found that the open-ended explanation condition did not show any significant differences in ordinary post-test problem-solving accuracy, suggesting that the intervention did not produce broad learning-gain advantages on practiced or similar problems. Second, we also found no significant condition differences on transfer answer selection, including NEI transfer MCQ accuracy, although the NEI MCQ estimate was in the same direction as the explanation-quality effect. Third, the open-ended condition produced significantly higher-quality explanations on NEI transfer problems ($p$\,=\,.030), with a marginally significant advantage on combined open-ended transfer measures ($p$\,=\,.057). Thus, the strongest evidence is not that open-ended self-explanation improved all forms of learning or transfer, but that it improved learners' ability to articulate reasoning on transfer problems. Notably, this occurred even though the open-ended condition showed similar observed learning gains to control while completing roughly one-quarter as many practice problems, a pattern consistent with the possibility that deeper processing per problem may compensate for reduced practice volume.

\noindent \textbf{Self-explanation under fixed practice time.} Prior work found that self-explanation advantages can disappear when control groups receive equivalent practice time \cite{matthews2009pursuit,mceldoon2013self}. Our 60-minute fixed practice session similarly showed no condition differences in overall learning gains, but it revealed a significant advantage on explanation quality under time-controlled conditions. Two differences may help explain this pattern. First, prior studies used unprompted verbal explanations without feedback on quality, whereas our open-ended condition provided adaptive LLM feedback that supported iterative revision. This interpretation is consistent with Aleven and Koedinger's \cite{aleven2001towards} finding that dialogue-based tutoring with corrective feedback can achieve equivalent learning with fewer problems. Second, the largest observed effect was concentrated on NEI problems requiring metacognitive judgment, whereas prior studies focused more on near transfer within the same problem type.

\noindent \textbf{Reasoning articulation vs.\ answer selection.} One interpretation is that open-ended self-explanation with LLM feedback trains more precise reasoning articulation, directly assessed by open-ended items but not captured by MCQ. Another possibility is that explanation quality reflects a deeper understanding that our MCQ items lacked the sensitivity to detect in this sample size (the MCQ effect was in the same direction, $d$\,=\,0.26). The effect on NEI problems may further reflect metacognitive demands: recognizing insufficient information requires monitoring the boundaries of one's knowledge \cite{chi1989self}, a skill that explaining one's reasoning may specifically cultivate \cite{renkl1997learning}.

\noindent \textbf{Benefit of LLM-powered self-explanation.} Traditional dialogue-based self-explanation tutors required hundreds of handcrafted rules \cite{aleven2001towards}. LLMs help address this bottleneck: with appropriate rubrics, they can evaluate diverse explanations without domain-specific rule engineering, achieving agreement with human consensus of $\kappa$\,=\,0.78. From an ICAP perspective \cite{chi2014icap}, this may also shift open-ended self-explanation beyond constructive toward interactive: learners generate an explanation, receive adaptive feedback, and revise their reasoning in response.

\noindent \textbf{Limitations.} The open-ended condition differed from the control not only in response format but also in feedback intensity: learners generated explanations, received iterative AI evaluation, and saw a reference explanation after repeated unsuccessful attempts. Thus, the present study cannot isolate whether the observed NEI explanation-quality advantage came from open-ended generation, adaptive feedback, the reference explanation, or their combination; future work should include feedback-only or limited-feedback controls. In addition, although the LLM grader showed substantial agreement with human ratings, we used one shared rubric and prompt across open-ended responses. This design supports consistency across items but may be less precise than item-specific rubrics that capture the particular reasoning required by each problem. Future work should develop more detailed item-level rubrics and validate them with additional human grading. Finally, our sample (N\,=\,92) may have been underpowered for smaller effects, and results from this introductory calculus unit may not generalize to other mathematical domains. The menu-based condition also differed from prior ITS implementations \cite{aleven2002effective}, where learners selected relevant principles from a larger set rather than choosing among one correct explanation and two misconceptions.

\bibliographystyle{splncs04}
\bibliography{reference}

\end{document}